%% file: cavityletter.tex
\documentclass[aps,prl,reprint,twocolumn,amsmath,amssymb,floatfix,showpacs,unsortedaddress,superscriptaddress]{revtex4-1}
\usepackage{graphicx}

\begin{document}

\title{Growing point-to-set length scale correlates with growing relaxation times in model supercooled liquids}

\author{Glen M. Hocky}
\affiliation{Department of Chemistry, Columbia University, 3000 Broadway, New York, New York 10027, USA}
\author{Thomas E. Markland}
\affiliation{Department of Chemistry, Stanford University, 333 Campus Drive, Stanford, California 94305, USA}
\author{David R. Reichman}
\email{drr2103@columbia.edu}
\affiliation{Department of Chemistry, Columbia University, 3000 Broadway, New York, New York 10027, USA}


\begin{abstract}
It has been demonstrated recently that supercooled liquids sharing simple structural features (e.g. pair distribution functions) may exhibit strikingly distinct dynamical behavior.
Here we show that a more subtle structural feature correlates with relaxation times in three simulated systems that have nearly identical radial distribution functions but starkly different dynamical behavior.
In particular, for the first time we determine the thermodynamic ``point-to-set'' length scale in several canonical model systems and demonstrate the quantitative connection between this length scale and the growth of relaxation times.
Our results provide clues necessary for distinguishing competing theories of the glass transition.
\end{abstract}

\maketitle

The search for growing length scales accompanying the glass transition has been a major focus of the field for two decades \cite{Berthier-RMP2011}.
Great progress has been made in quantifying the behavior of dynamical length scales associated with emergent dynamical heterogeneity, such as $\xi_4$, the length scale of the 4-point susceptibility \cite{Franz-PhilMagB1999}.
Less mature is our understanding of possible thermodynamic length scales that grow upon supercooling.  In some theories of the glass transition, such as the random first-order theory (RFOT), dynamical arrest is connected to a particular thermodynamically-based length scale associated with the depletion of independent particle configurations constrained by neighboring particles \cite{Kirkpatrick-PRA1989,Lubchenko-ARPC2007}.
Within this viewpoint the putative structural length scale is distinct from $\xi_4$ for modest supercooling \cite{Franz-JPhysA2007}.
In other approaches this length scale may be identical to that associated with dynamical heterogeneity at all temperatures \cite{Chandler-ARPC2010,Jack-JCP2005}.
Furthermore, in these contrasting viewpoints the relationship of the various growing length scales with growing relaxation times are different.
Clearly, elevating our level of knowledge of growing thermodynamic length scales associated with glass formation to that of dynamical ones is of paramount importance in the continued quest for a deeper understanding of the behavior of supercooled liquids and glasses.

A precise definition of one type of non-trivial structural length scale akin to that envisioned in theories like RFOT was put forward by Bouchaud and Biroli in 2004 \cite{Bouchaud-JCP2004}. Their work suggests a procedure for the extraction of this length scale in computer simulations.
This length is a measure of the distance scale over which particles are self-consistently pinned by other particles in their vicinity. The static correlations embodied by this length scale are known as ``point-to-set'' (PTS) correlations. In what follows, we will use $\xi_{PTS}$ to denote the PTS length scale.
There has been a recent flurry of activity in the extraction of variants of $\xi_{PTS}$ in a host of model glass forming systems \cite{Karmakar-PNAS2009,Jack-JCP2005,Biroli-NatPhys2008,Berthier-PhysRevE2012,Karmakar-PhysicaA2012,Cammarota-Arxiv2011,Charbonneau-PRL2012}.
In this work we use this technique to address the crucial question of correlation between $\xi_{PTS}$ and dynamics in systems with identical simple structural features (e.g. radial distribution functions, $g(r)$) but noticeably distinct relaxation time scales.

\begin{figure}[t]
\centering
\includegraphics[scale=1.0]{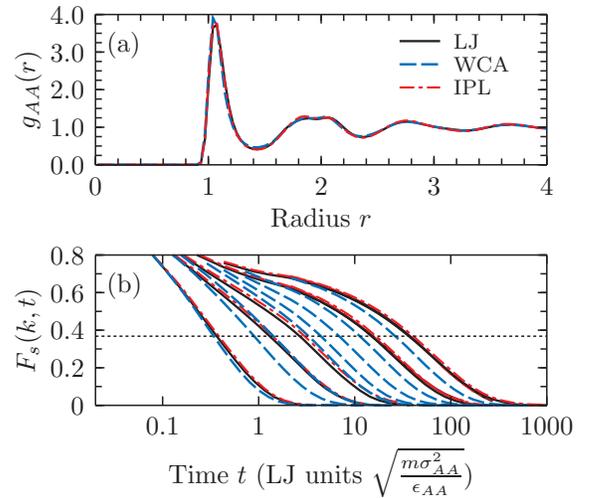}
\caption{
{\bf (a)} The radial distribution functions for A-type particles at $T=0.6$ for the three models studied as an illustration of the structural similarity between the models. Other temperatures and pairs are shown in the supplemental information (SI).
{\bf (b)} The self-intermediate scattering functions for $A$-type particles at $k=7.25$ for NVE molecular dynamics simulations of the bulk systems. The alpha relaxation, $\tau_\alpha$, is defined as the $1/e$ time of $F_s(k,t)$, illustrated by the horizontal dotted line.
Curves shown for all models are, from L-R, $T=\{2.0,1.0,0.8,0.6,0.55\}$ and additionally for WCA $T=\{0.5,0.45,0.4\}$.
Note the LJ and IPL match at all temperatures while the WCA matches only at T=2.0 and decays significantly faster at lower temperatures.
These temperature represent the regime computationally accessible to our cavity simulations. Parameters for the models can be found in the SI.
\label{fig:sd}
}
\end{figure}

We consider three closely related systems, chosen specifically to address this problem. The first two are the standard Kob-Andersen binary Lennard-Jones mixture (LJ) \cite{KALJ-I-PRE1995} and its Weeks-Chandler-Andersen truncation (WCA) \cite{Weeks-JCP1971}, which were previously found to have significantly different dynamical behavior at supercooled temperatures \cite{BerthierTarjus-PRL2009}, despite having nearly identical two-body static correlations at all temperatures (e.g. measured by $g(r)$). The third is a system characterized by a repulsive inverse power law potential (IPL) and was constructed based on the composition and parameters of the LJ system to reproduce both two-body structural features and the dynamics of the LJ model \cite{Pedersen-PRL2010}. The similarities in pair correlations are illustrated in Fig.~\ref{fig:sd}(a), and the structural relaxation times for the temperatures in this study are shown in Fig.~\ref{fig:sd}(b). While it should be noted that minor differences exist in $g(r)$ between the three systems \cite{Pedersen-PRL2010} (see supplemental information (SI)), they are far too small to account for the significantly weaker temperature dependence of relaxation times in the WCA system. In particular, neither mode-coupling theory nor activation-based theories that rely solely on the radial distribution function can account for this distinction \cite{Berthier-EPJE2011}. These three systems will allow us to study whether $\xi_{PTS}$ is sensitive to structural differences in the three models not appearing at the level of $g(r)$, and if so whether $\xi_{PTS}$ can distinguish the LJ and IPL from the WCA.

To measure $\xi_{PTS}$, we follow the protocol of Ref.~\onlinecite{Biroli-NatPhys2008}. All simulations were performed by the Monte Carlo (MC) technique of Ref.~\onlinecite{Grigera-PRE2001}. Bulk equilibrium configurations are generated at a desired temperature, and then cavities are constructed by freezing the particles outside a sphere of radius $R$. The center of the cavity is partitioned into $\tilde{N}$ cubes of side length $l$.
The overlap is then defined as $q(R,t)=(l^3 \tilde{N})^{-1} \sum_{i=1}^{\tilde{N}} \langle n_i(t_0)n_i(t_0+t) \rangle$ 
where brackets denote both a thermodynamic average and an average over independent cavities, and $n_i(t)$ is a binary digit specifying whether a particle is in box $i$ at ``time'' $t$ (time here standing for any measure of simulation progress). As the cavity evolves in simulation, this quantity decays to a plateau which is independent of time and denoted $q(R)= \lim_{t \rightarrow \infty} q(R,t)$.  Overlaps generated in this fashion will be referred to as ``standard overlaps''. With the preceding definition, two independent configurations will have an overlap $q=\rho l^3$ which is also the value of $q(R\rightarrow\infty).$ We will henceforth subtract off this bulk overlap from $q$, and denote the resulting value $\tilde{q}$. Further details can be found in the SI.

\begin{figure}[t]
\centering
\includegraphics[scale=1.0]{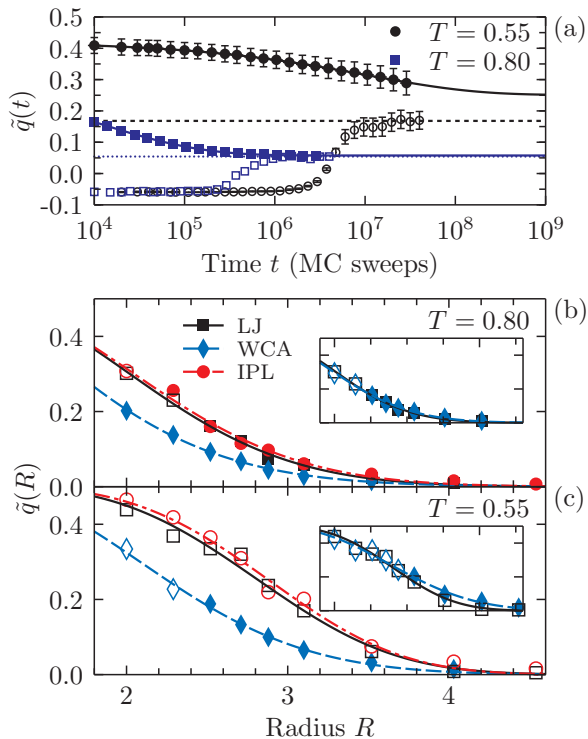}
\caption{ {\bf (a)} 
	Overlap as a function of MC sweeps is shown for cavity size $R=3.1$ at two different temperatures.
        Filled symbols show the overlap from MC dynamics and solid lines show stretched exponential fits to this data. 
	Open symbols show the overlap from PSA, and dashed and dotted lines show horizontal fits to the long time plateau.
        Note that the overlap from PSA and MC dynamics do not meet at the lower temperature. 
        In PSA, the particles have reduced diameters and may sample many more configurations than when they are full-sized, thus the PSA data go below the bulk value at short times.
        Error bars are from a bootstrap analysis and for the higher temperature are substantially smaller than the symbols shown.
	{\bf (b)}, {\bf (c)} Overlap as a function of cavity size at two different temperatures.
	Here, closed symbols show standard overlaps, open symbols overlaps from PSA.
        Lines through the data are compressed exponential fits with the form discussed in the text.
	The LJ system is represented by solid lines, the WCA dashed, and the IPL dash-dotted.
	The insets show the same LJ data from the main figure with the same axis limits.
	Overlaid are data from the WCA system at which the length is most similar --- $T=0.6$ in (b) and $T=0.4$ in (c). Bootstrap errors are, at largest, the size of the symbols shown.}
\label{fig:overlap}
\end{figure}

A major complication of the algorithm sketched above is that the plateau value $q(R)$ will over-estimate the true thermodynamic value of the overlap if the particle configuration inside the cavity breaks ergodicity or if the confinement simply induces relaxation on a time scale beyond that accessible to our simulations. We note that for all three systems studied here, relaxation times increase dramatically as cavity radius is decreased.  The technique of particle swapping \cite{Grigera-PRE2001} ameliorates this problem in some systems, but is not effective in the systems studied here, as swap moves that exchange particles of different species are almost never accepted at supercooled temperatures.
To test for convergence to the true (thermodynamic) value of $\tilde{q}(R)$, Cavagna and coworkers have proposed a technique based on the insertion of a {\em random} configuration of the same particles in the same cavity \cite{Cavagna-PRL2007}. Initially such a configuration will have, on average, zero overlap ($\tilde{q}=0$) with the pre-randomized configuration, but the structure of the boundary will induce finite overlap at long times. If this value measured with respect to the initial configuration yields the same value of $\tilde{q}(R)$ as that extracted from the direct decay of $\tilde{q}(R,t)$,  one can be confident that the true thermodynamic value of the overlap has been obtained \cite{Cavagna-PRL2007}. This quantity is difficult to converge in our systems, hence we have implemented an approach which we call ``particle size annealing'' (PSA).  In this method the particles inside the cavity are instantaneously reduced in size such that their positions quickly randomize, and are then evolved with the constraint of the cavity in place while their diameters are slowly tuned back to their original size. 
We have found this method, which is similar in spirit to algorithms used to generate randomly jammed packings of hard spheres \cite{Lubachevsky-JStatPhys1990}, is more efficient and reliable in the generation of converged overlap values than standard Monte Carlo sampling.  In what follows, we use PSA both as a check of the convergence of $\tilde{q}(R)$ to its thermodynamic value and as a means of generating estimates of $\tilde{q}(R)$ for small $R$ values.  Specifically, for radii where standard and PSA overlap values match within error bars, the standard overlap is taken as the thermodynamic overlap.  For smaller radii, the value from PSA yields a lower bound to the converged thermodynamic value of $\tilde{q}(R)$. Further, we expect PSA to yield estimates of $\tilde{q}(R)$ that are extremely close to the desired thermodynamic values. This expectation arises from a comparison with other sampling techniques (e.g. replica exchange) and direct Monte Carlo sampling. A discussion of these comparisons will be made in a future publication. Details of our approach may be found in the SI. Example results for a single cavity size can be seen in Fig.~\ref{fig:overlap}(a).

We now turn to a discussion of the extraction of $\xi_{PTS}$ from the spatial decay of the converged overlap function $\tilde{q}(R)$. Overlaps were fit to a generalized compressed exponential of the form $\tilde{q}(R)=A\exp(-\left (\frac{R-a}{\xi_{PTS}} \right )^\eta)$. Though a previous study used a pure compressed exponential (i.e. $a=0$)\cite{Biroli-NatPhys2008}, we choose $a=1$, physically motivated by the fact that cavities with $R\sim1$ should on average contain a single particle and that the overlap properties at this cavity size should not be sensitive to growing amorphous order. Furthermore we do not expect the same compressed exponential form to extend to cavities containing on average fewer than one particle. Fixing an $a>0$ allows us to perform a two-parameter fit with fixed $A$, leading to values of $\xi_{PTS}$ with much smaller statistical variance. We find that for $a=1$, $A=0.5$ gives good fits to the data for all three systems at all temperatures studied; some example fits can be seen in Fig.~\ref{fig:overlap}(b) and \ref{fig:overlap}(c). An extended discussion of our fitting choices and methodology can be found in the SI, as well as a table of fit parameters extracted from the data. 

\begin{figure}[t]
\centering
\includegraphics[scale=1.0]{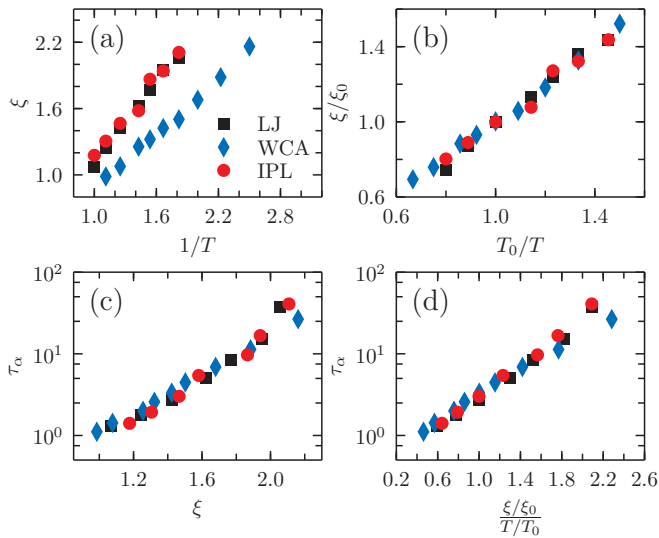}
	\caption{ {\bf (a) } Length vs. inverse temperature. For clarity, $\xi$ here represents $\xi_{PTS}$ of the text. A clear distinction of scale between the WCA system and the other two systems should be noted.
	{\bf (b) } The same data as in (a) but scaled to a temperature where all three models have the same length, $T_0=0.8$ for the LJ and IPL and $T_0=0.6$ for the WCA. Here the length $\xi_0$ is $\approx 1.4$.
	{\bf (c) } Structural relaxation time $\tau_\alpha$ vs. length $\xi$. The length scale correlates directly with the relaxation time scale.
	{\bf (d) } Structural relaxation time vs. reduced length over reduced temperature. $T_0$ and $\xi_0$ are the values giving collapse in (b). 
        Error bars from our bootstrap analysis are approximately the size of the data symbols. All data are reported in the SI.
}
\label{fig:lengths}
\end{figure}

Fig.~\ref{fig:lengths}(a) illustrates our first main result, namely the growth of the absolute thermodynamic length scale $\xi_{PTS}$ for all three systems discussed above.  Several notable features deserve mention.  First, it is clear that the length scale $\xi_{PTS}$ grows unambiguously as temperature is lowered.  This is fully consistent with other recent studies that demonstrate growth of $\xi_{PTS}$ in a variety of pinning geometries \cite{Karmakar-PNAS2009,Jack-JCP2005,Biroli-NatPhys2008,Berthier-PhysRevE2012,Karmakar-PhysicaA2012,Cammarota-Arxiv2011,Charbonneau-PRL2012}. Second, the distinction between the magnitude of $\xi_{PTS}$ in the WCA system compared to the LJ and IPL systems at the same absolute temperature is stark.  Despite the fact that the pair distribution functions of all three systems are nearly identical, the more subtle structural marker $\xi_{PTS}$ can clearly distinguish the WCA system  from the other two. The lengths of the IPL and LJ are found to be nearly identical at all temperatures and much larger than those found for the WCA. 

We now address the crucial question of correlation with relaxation times in a quantitative manner.  Is $\xi_{PTS}$ correlated in a one-to-one manner with the alpha relaxation time $\tau_\alpha$ extracted from the self-intermediate scattering functions of the systems under investigation?  A key component of the answer to this question may be found in Fig.~\ref{fig:lengths}(b).  While the absolute magnitude of $\xi_{PTS}$ in the WCA system is clearly smaller than that of the other two systems at the same temperature, the lengths of all three systems collapse when temperature is scaled to the value where $\xi_{PTS}\approx1.4$. These temperatures are quite similar to values obtained for ``onset`` temperatures obtained by independent means in earlier work \cite{BerthierTarjus-PRL2009}. 
In Fig.~\ref{fig:lengths}(c) we show the behavior of relaxation times $\tau_{\alpha}$ (obtained from NVE molecular dynamics of bulk equilibrium systems) as a function of $\xi_{PTS}$ in all three systems.  Reasonable data collapse is found, signifying a strong correlation between these quantities in all three systems.  Fig.~\ref{fig:lengths}(d) shows a scaling plot where the independent variable takes an activated form $\ln(\tau_{\alpha}) \sim \xi_{PTS}/T$, normalized by the values of the lengths and temperatures found from Fig.~\ref{fig:lengths}(b). The degree of collapse is similar to that shown in Fig.~\ref{fig:lengths}(c) with no adjustable parameters, making it difficult to distinguish between scaling forms where the temperature dependence of the growth of $\tau_{\alpha}$ may be attributed purely to the growth of $\xi_{PTS}$ alone or the activated form commonly found in the literature \cite{Berthier-RMP2011,Kirkpatrick-PRA1989,Lubchenko-ARPC2007}. Furthermore, both Figs.~\ref{fig:lengths}(c) and \ref{fig:lengths}(d) show slight but systematic deviations from perfect data collapse, suggesting that similar but distinct exponents associated with the scaling variable ($\xi_{PTS}$ or $\xi_{PTS}/T$) would be required to extend these scaling plots to larger $\tau_\alpha$ values; this may be similar to what is seen in Ref.~\onlinecite{Coslovich-PRE2011} \footnote{In Fig.~8 of Ref.~\cite{Coslovich-PRE2011}, Coslovich shows a relationship between cluster size of locally preferred structures in the LJ and WCA systems. Though he finds deviations at low temperatures, his Fig.~8 also implies approximate scaling of the ``length'' scale $\sim N_{LPS}^{1/3}$ with $\tau_\alpha$ in the temperature regime of our work. A relationship between such clusters and $\xi_{PTS}$ is not yet established.}. Regardless, the correlation between $\xi_{PTS}$ and relaxation time growth is striking. Finally, we note that within the resolution of our current data, when relaxation times of these three models are similar, it is not only $\xi_{PTS}$ as defined here that matches, but rather the full overlap profile $\tilde{q}(R)$; this is illustrated in the insets of Fig.~\ref{fig:overlap}. Hence, though our definition of $\xi_{PTS}$ is not unique, we expect any reasonable definition to give the same qualitative results as seen in Fig.~\ref{fig:lengths}.

The results presented here provide useful clues to the underlying causes of the viscous slowdown of dynamics during the supercooling process.
$\xi_{PTS}$ is the central structural feature associated with slow dynamics in RFOT-like theories \cite{Berthier-RMP2011,Kirkpatrick-PRA1989,Lubchenko-ARPC2007}.
In this sense, the results presented in Fig.~\ref{fig:lengths} would seem to be in harmony with that viewpoint. It should be noted, however, that within the standard RFOT, the lengths in the relatively high temperature regime are not expected to necessarily correlate strongly with relaxation times, as they do here.
Our results bear a resemblance to some features found in some short-ranged p-spin models, which serve as a possible paradigm for how the mean-field models that RFOT is based on are altered by fluctuations.
In particular, a weakly growing static length scale that is linearly correlated with $T\ln(\tau)$ all the way up through the high temperature regime as in Fig.~\ref{fig:lengths}(d) are features found in the model of Ref.~\onlinecite{Alvarez-PRB1996}.
Unfortunately, the static correlations in this model are spin-glass correlations, which are absent in real liquids. In general, studies such as those carried out here should shed light on how fluctuations modify (or destroy) the mean field behavior upon which the standard version of RFOT is based \cite{Moore-PRL2006,Cammarota-PRL2011}.

The mere existence of a growing $\xi_{PTS}$ is {\em not} in contradiction with a picture based on kinetically constrained models; however there are several aspects of our results which seem difficult to reconcile with the facilitated picture. To address this, one has to examine results from thermodynamically interacting models which can be mapped onto kinetically constrained models.
It has already been noted that local dynamics inside compact cavities show markedly increasing relaxation times as the cavity radius is decreased \cite{Berthier-PhysRevE2012}.
While the same behavior holds in the three systems studied here (not shown), the opposite is true for large cavities in the triangular plaquette model, the one model known to be dual to a kinetically constrained model with hierarchical behavior \cite{Jack-JCP2005}.
In addition, the absolute length scales found here are quite small.
Since $\xi_{PTS}$ might loosely be interpreted as the mean distance between defects in a facilitated model, this fact would seemingly translate into an unrealistically high density of defects with a rather weak temperature dependence.
Indeed, the magnitude of $\xi_{PTS}$ found in this work seems more consistent with the spatial extent of the ``defects'' found in a recent study than the separation between these regions \cite{Keys-PRX2011}.
Lastly, our results seem consistent with the idea that $\xi_{PTS}$ and $\xi_4$ are distinct in the regime of weak to moderate supercooling \cite{Karmakar-PNAS2009}, which would be unexpected in models such as the triangular plaquette model \cite{Franz-JPhysA2007,Jack-JCP2005} \footnote{This behavior is consistent with the square plaquette model, which is dual to the two-dimensional Fredrickson-Andersen model. This model has Arrhenius transport, unlike the atomistic systems studied here.}. It is important to bear in mind that plaquette models provide merely a small set of all possible mappings of an interacting system to one of kinetically constrained but thermodynamically non-interacting defects. It is unlikely, however, that any such mapping would have much to say about the small length scales found in this work, which are expected to be similar to the coarse graining length scale in any mapping to a kinetically constrained model \cite{Garrahan-PNAS2003}.

In conclusion, $\xi_{PTS}$ in three model glass forming systems has been been extracted.
Care has been exercised to avoid dynamical contamination of the overlap function.
We find that $\xi_{PTS}$ is small and grows systematically albeit modestly as temperature is lowered.
The spatial range of the thermodynamic overlap functions and extracted $\xi_{PTS}$ are clear discriminators of dynamical behavior, even when simple structural features such as pair distribution functions are blind to differences in relaxation times.
We find that in the cases studied here, $\xi_{PTS}$ correlates reasonably well in absolute terms with the relaxation times in all three systems.
Future work will be devoted to assessing if these correlations robustly hold over a wider range of temperatures and systems,
as well as connecting the results uncovered here to other recent work on non-trivial structural and thermodynamic markers of glassy behavior \cite{Coslovich-PRE2011,Karmakar-PNAS2009,Karmakar-PhysicaA2012}.

\begin{acknowledgments}
This research was performed on the following computing resources provided by the National Science Foundation (NSF), National Institutes of Health (NIH), and Depart of Energy (DOE) Office of Science: the Open Science Grid (supported by the NSF and DOE) and EngageVO (NSF Grant No. OCI-0753335), the PADS (NSF Grant No. OCI-0821678) and Beagle (NIH Grant No. S10-RR029030-01) systems at the Computation Institute, a joint institute of Argonne National Laboratory and the University of Chicago, and the Extreme Science and Engineering Discovery Environment (XSEDE, NSF Grant No. OCI-1053575). We thank Michael Wilde and the Swift development team for discussions and advice on using the Swift parallel scripting language \cite{Wilde-Parallel2011} to facilitate the very large number of simulations that were performed on this diverse set of resources for this investigation. Their assistance was supported in part by NSF Grant No. OCI-1007115. G.M.H., T.E.M., and D.R.R. were supported by the NSF through a Graduate Research Fellowship (Grant No. DGE-07-07425), Grant No. CHE-0910943, and Grant No. CHE-0719089, respectively. We thank Bruce J. Berne, Ludovic Berthier, Giulio Biroli, Chiara Cammarota, Andrea Cavagna, Daniele Coslovich,  Tom\`{a}s Grigera, Robert Jack, Walter Kob, and Paolo Verrocchio for helpful conversations.
\end{acknowledgments}

\bibliographystyle{apsrev}
\bibliography{cavityletter}

\clearpage
\section{Supplemental information}
\input{cavityletter_supplement_text.txt}

\end{document}

%% file: cavityletter_supplement_text.txt
\subsection{System Details}
The LJ model is an 80:20 binary mixture of Lennard-Jones spheres with size and interaction parameters given by $\sigma_{AB}/\sigma_{AA}=0.8$, $\sigma_{BB}/\sigma_{AA}=0.88$, and $\epsilon_{AB}/\epsilon_{AA}=1.5$, $\epsilon_{BB}/\epsilon_{AA} = 0.5$ \cite{KALJ-I-PRE1995}.
The WCA system is characterized in the same way, but the interaction potential is shifted up by $\epsilon_{ij}$ and cut off at $r=2^{1/6} \sigma_{ij}$ \cite{Weeks-JCP1971}.
The IPL system is characterized by an inverse-power law potential given by $V(r_{ij})=A\epsilon_{ij}(\sigma_{ij}/r_{ij})^n$ with $n=15.48$ and $A=1.945$ and the same size and energy parameters as the LJ system.  All quantities are reported in standard reduced Lennard-Jones units, and all systems are studied at number density $\rho=1.2$ \cite{Pedersen-PRL2010}.  

\begin{figure*}
\centering
\includegraphics[scale=1.0]{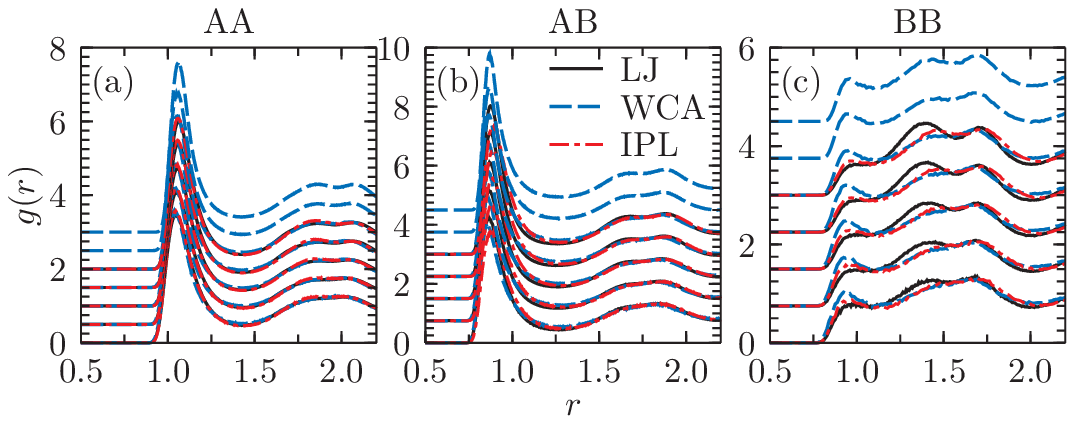}
\caption{Radial distribution function for the three models. Shown are temperatures $T=\{0.9,0.8,0.7,0.6,0.55\}$ for all models and also $T=\{0.5,0.4\}$ for the WCA model. Lines are shifted for clarity by 0.5 in panel (a) and by $0.75$ in panels (b) and (c). The highest temperatures are at the bottom. \label{fig:rdfs}
}
\end{figure*}

Radial distribution functions have been generated for the three models at the temperatures studied in the main text. A figure showing these correlations for most of the temperatures in the relevant temperature regime can be found in Fig.~\ref{fig:rdfs}. Though differences do exist between the $g(r)$'s at all temperatures, the differences stay relatively constant and are small. We do not attempt to demonstrate that these differences are too small to account for dynamical differences observed between the models, as this has already been discussed in prior work \cite{BerthierTarjus-PRL2009}.

\subsection{Simulation Details}
Eight independent configurations of 4050 particles were equilibrated at each temperature of interest for each system using Swap-MC \cite{Grigera-PRE2001}.
Six locations were chosen randomly in each of the eight independent configurations and particles whose centers fell outside a sphere of a given radius were frozen.
This was done with the condition that the ratio of particle types and the density within a sphere of a given size matched that of the bulk.
In order to ensure this, radii were defined by the total number of particles inside that spherical cavity i.e. $R=(\frac{3N}{4\pi\rho})^{1/3}$ with $N$ the number of particles and $\rho$ the bulk density.
Moves that took a particle outside of the cavity were rejected (although without this restriction, particles escaped only rarely and usually returned immediately).
The overlap (averaged over the 48 independent cavities) was monitored as a function of time until reaching a plateau.
This plateau value was extracted through a fit to a shifted stretched exponential and the long time limit was taken as the thermodynamic overlap.
Error in this value was estimated by bootstrapping a distribution of curves to fit \cite{Efron1993-Bootstrap}. 

As in Ref.~\onlinecite{Biroli-NatPhys2008}, we use 125 boxes (a 5x5x5 cube) to probe the overlap. In Ref.~\onlinecite{Biroli-NatPhys2008}, the authors choose $l^3=0.062876$ which they found was small enough to prevent two particles from appearing in the same box at number density $\rho=1$. Our systems studied were at $\rho=1.2$ so we chose to use box sizes approximately 1.2 times smaller, or $l^3=0.05$. The bulk overlap for our systems is therefore $\rho l^3=0.06$.

\subsection{PSA Details}
In the Particle Size Annealing (PSA) method, the interaction between particles is adjusted such that $\sigma_{ij} \rightarrow \frac{\lambda_i \sigma_{ij}+ \lambda_j \sigma_{ij}}{2} = \frac{\lambda_i+\lambda_j}{2}\sigma_{ij}$.
Each fixed particle outside of the cavity is assigned $\lambda_i=1$ and each cavity particle is assigned a $\lambda_i$ based on some tunable switching function.
The switching function we employ is given by $\lambda(t)=\lambda_0+(1-\lambda_0)\left[1 - (1-(t/\tau_s) )^\gamma \right ]$ where $\lambda_0$ is the smallest particle size, $\tau_s$ is the time over which switching occurs and $\gamma$ controls the steepness of the curve.
For our purposes, we empirically chose to run at $\lambda_0=0.6$ for several bulk $\tau_\alpha$, and then to anneal with $\gamma=4$ and $\tau_s$ ten times longer than the length of the randomization phase.
These simulations are run using the Swap-MC algorithm because for $\lambda(t)<1$, many swap moves are accepted and this greatly enhances sampling. 
Because these data are noisier than the thermodynamically averaged overlap from an equilibrium simulation (due to the fact that only the 48 initial cavity configurations exist for reference), two PSA simulations were run per cavity. 
These data were logarithmically binned and averaged across all of the cavities for a given system and temperature. 
The cavity overlap was taken by fitting the plateau of this curve to a horizontal line.

\subsection{Fits and Extraction of a Length}
Here we comment on the fitting form chosen for extracting a length from the overlap data and report the resulting fit parameters. The authors of Ref.~\onlinecite{Biroli-NatPhys2008} find that a compressed exponential form $\tilde{q}(R)=A\exp(-\left (\frac{R}{\xi_{PTS}} \right )^\eta)$ fits their low temperature data well. However, they fix $\eta=1$ to get good fits to their higher temperature data, resulting in a sharp jump in the lengths extracted. Physically, we do not expect their fitting form (rather, any fitting form) to hold for all $R>0$ given that the overlap quantity is not well defined for very small cavities. We expect that when $R=1$, i.e. the cavity radius is approximately the size of a large particle "diameter", to have a cavity with about one particle on average. For a cavity of this size, we would expect an overlap between zero and one, and moreover we predict this value to be approximately temperature independent and devoid of information related to emergent amorphous order. By still fitting our data to a compressed exponential, but shifting it to start at $R=1$, it was found that the form $\tilde{q}(R)=A\exp(-\left (\frac{R-a}{\xi_{PTS}} \right )^\eta)$ with $a=1$ fits all of the data at all temperatures given a fixed $A$ value in the range $(0.45-0.6)$. Thus a two parameter fit to all of the data becomes viable. By sweeping the value of $A$ in this range, it was found that $A=0.5$ fits the data best across the three models, though slightly better fits could be done by picking a different $A$ for each model. We chose to use the same $A$ value for all three models, because we felt this resulted in a value of $\xi$ that was more equivalent when comparing the models and because this decreases the possibility of spurious overfitting. From an alternative perspective one sees by plugging in $\tilde{q}(\xi+1)=0.5/e$ that our fit form is reporting as a length the cavity size where $\tilde{q}(R+1)\approx 0.18$ (again motivated by the physical argument that nontrivial amorphous order sets in at scales $R>1$). Hence this definition could be used independently of fit form if an alternative function were determined to fit the data better. 

\begin{figure}
\centering
\includegraphics[scale=1.0]{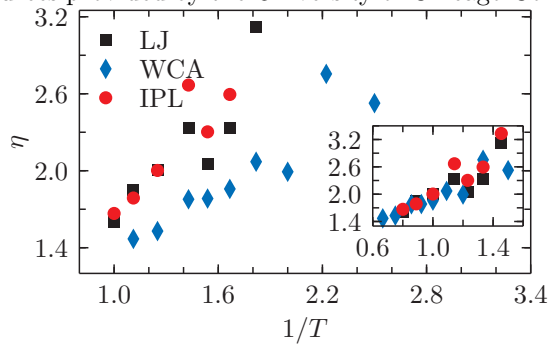}
\caption{Fit parameter $\eta$ as a function of inverse temperature. Inset: $\eta$ plotted against scaled inverse temperature where the scaling is as in Fig.~3(b). \label{fig:eta}
}
\end{figure}

Given this flexibility in method for extracting the length, we do not attempt to ascribe meaning to the fit parameter $\eta$ or to the exact magnitude of $\xi_{PTS}$. Nevertheless, we show in Fig.~\ref{fig:eta} that generally the same qualitative conclusions can be drawn for $\eta$ as for $\xi_{PTS}$, namely that the $\eta$ values can distinguish the LJ and IPL from the WCA at a given temperature but collapse when scaled to the onset temperatures. However, there is much more noise in the $\eta$ values as extracted. As stated earlier, the errors in the fit parameters were determined by generating a bootstrap distribution of $\tilde{q}(R)$ curves from the 48 standard $\tilde{q}(R,t)$ or 96 PSA $\tilde{q}(R,t)$ data and calculating the standard deviation of that distribution \cite{Efron1993-Bootstrap}.  

We report the fit values obtained using the methods discussed in this section in Tab.~\ref{tab:lengths}. 

\begin{table*}
\begin{tabular}{ | c || c | c || c | c || c | c | }
\hline
& LJ & & IPL & & WCA & \\
\hline
Temp & $\xi$ & $\eta$ & $\xi$ & $\eta$ & $\xi$ & $\eta$ \\
\hline
1.0  & $1.067 \pm 0.005$ & $1.599 \pm 0.019$ & $1.177 \pm 0.003$ & $1.667 \pm 0.012$ &                   &                   \\
0.9  & $1.243 \pm 0.002$ & $1.853 \pm 0.009$ & $1.305 \pm 0.002$ & $1.787 \pm 0.010$ & $0.986 \pm 0.003$ & $1.469 \pm 0.010$ \\
0.8  & $1.427 \pm 0.002$ & $2.008 \pm 0.014$ & $1.466 \pm 0.001$ & $2.002 \pm 0.000$ & $1.077 \pm 0.002$ & $1.530 \pm 0.006$ \\
0.7  & $1.620 \pm 0.002$ & $2.331 \pm 0.014$ & $1.581 \pm 0.119$ & $2.668 \pm 0.984$ & $1.255 \pm 0.003$ & $1.778 \pm 0.011$ \\
0.65 & $1.770 \pm 0.003$ & $2.050 \pm 0.013$ & $1.865 \pm 0.002$ & $2.303 \pm 0.013$ & $1.322 \pm 0.002$ & $1.783 \pm 0.007$ \\
0.6  & $1.949 \pm 0.002$ & $2.329 \pm 0.011$ & $1.940 \pm 0.003$ & $2.594 \pm 0.014$ & $1.423 \pm 0.003$ & $1.858 \pm 0.011$ \\
0.55 & $2.055 \pm 0.002$ & $3.119 \pm 0.017$ & $2.108 \pm 0.002$ & $3.332 \pm 0.016$ & $1.502 \pm 0.003$ & $2.070 \pm 0.015$ \\
0.5  &                   &                   &                   &                   & $1.679 \pm 0.002$ & $1.992 \pm 0.001$ \\
0.45 &                   &                   &                   &                   & $1.884 \pm 0.002$ & $2.755 \pm 0.014$ \\
0.4  &                   &                   &                   &                   & $2.161 \pm 0.003$ & $2.527 \pm 0.013$ \\
\hline
\end{tabular}
\caption{Values extracted from fitting as a function of temperature $\tilde{q}(R)=0.5 \exp{\left(\frac{R-1}{\xi}\right )^\eta}$ for the three systems studied. Errors are calculated as discussed in the text of the supplemental information.\label{tab:lengths}
}
\end{table*}

\subsection{Computational Details}
The computations above comprised approximately 30000 independent tasks broken into hundreds of thousands of jobs, totaling more than two million hours of computational time. They were executed in parallel on resources provided by the University of Chicago Computation Institute, the Open Science Grid, and the Extreme Science and Engineering Discovery Environment. These resources were leveraged by expressing the simulations in the Swift parallel scripting language \cite{Wilde-Parallel2011}. Example Swift code for this project can be found in Ref.~\onlinecite{Wilde-Parallel2011}.